\documentclass[12pt]{article}
\usepackage[dvips]{graphicx}
\usepackage[dvips]{epsfig}

\usepackage[T2A]{fontenc}
\usepackage[cp1251]{inputenc}
\usepackage[russian]{babel}

\begin{document}

\begin{center}
{\large\textbf{\textsf{Fractional Models of Cosmic Ray
Acceleration in the Galaxy }}}\footnote{The work is supported by
RFBR, grant 10-01-00608}
\end{center}
\begin{center}
\textbf{\textsf{}}$^{}$
\end{center}
\begin{center}
{\sl V. V. Uchaikin}
\end{center}
\begin{center}
Ulyanovsk State University, 432970 Ulyanovsk, L. Tolstoy 42
\end{center}
\begin{center}
vuchaikin@gmail.com
\end{center}
\begin{abstract}
Possible formulations of the problem of cosmic rays acceleration
in the interstellar galactic medium are considered with the use of
fractional differential equations. The applied technique has been
physically justified. A Fermi result has been generalized to the
case of the acceleration of particles in shock waves in the
supernovae remnants fractally distributed in the Galaxy.
\end{abstract}


\section{Introduction}
The physical meaning of the fractional differential equation
$$
\left[_0\textsf{D}_t^\beta+D(-\triangle)^{\alpha/2}\right]
\psi(\mathbf{x},t)=\delta(\mathbf{x})\delta_\beta(t),\  \delta_\beta(t)
\equiv t_+^{-\beta}/\Gamma(1-\beta),\ \alpha\in(0,2],\ \beta\in(0,1],\eqno(1)
$$
was discussed in previous work [1] in the continuous time random
walk (CTRW) model in view of the problems arising when it is
applied to the description of the transport of cosmic rays in the
Galaxy. In this work, the CTRW approach is considered in
application to the description of the acceleration (more
precisely, reacceleration) of cosmic rays. In contrast to the
preceding problem, this concerns the behavior of particles in the
momentum space. Successive interactions (collisions) of a charged
particle with more or less localized inhomogeneities of the
magnetic field from slowly moving magnetic clouds mentioned by
Fermi in his pioneering works [2] to strong shock waves in the
supernovae remnants mentioned by Berezhko and Krymskii in review
[3] can be considered as instantaneous jumps from one point of the
momentum space to another one. The momenta $\Delta \mathbf{p}_i$
acquired by the particle in these collisions are random and, even
for their isotropic distribution, the point
$$\mathbf{p}=\mathbf{p}_0+\Delta\mathbf{p}_1+\Delta
\mathbf{p}_2+\Delta\mathbf{p}_3+\dots,$$ representing the particle
in the momentum space moves away from the point (momentum) of the
acceleration injection $\mathbf{p}_0$ similar to a Brownian
particle; this behavior means the further acceleration
(reacceleration) of the particle. However, only a certain fraction
of the particles moving away from the center are accelerated. This
fluctuation component of the mechanism of the acceleration of
cosmic rays is analyzed in this work.

From the statistical point of view, the main consequence of the
Fermi conclusion is that the exponential increase in the energy of
the accelerated particle with time, $E = E_0 e^{at}$, and the
exponential distribution of the age of the detected particles,
$d\textsf{P}=\exp(-t/\tau)dt/\tau$, are sufficient for the
formation of the power-law energy spectrum $N(E)$. That is all.
There are no other sources of fluctuations taken into account in
the Fermi model. This result is evident: What fluctuations can
else exist if an increase in the energy by a factor of $e$
requires $10^8$ collisions according to the Fermi estimation? A
more significant effect could be produced by more rare acts with
more large acceleration in each of them. Among these processes are
the aforementioned interactions with strong shock waves, when even
a single interaction can increase the energy of the particle by a
factor of $7\div13$ (see [4, p. 449]). To this end, it is
appropriate to pass from the degenerate spectral function
$\delta(E-E_0 e^{at})$ characterizing the determinate process of
the Fermi acceleration to the continuous function $n(E, t)$
related to the momentum distribution $f(\mathbf{p}, t)$ as

$$
n(E,t)=\int\limits\delta(E-E(\mathbf{p}))f(\mathbf{p},t)d\mathbf{p}.
$$
Under the Fermi assumption that the parameter $\tau$ is
independent of the energy, the desired spectrum can be represented
in the form
$$
N(E)\equiv N(E;\tau)=\tau^{-1}\left[\int\limits_0^\infty n(E,t)
e^{-t/\tau}dt\right]=\tau^{-1}\widehat{n}(E,\tau^{-1}),\eqno(3)
$$
where
$$
\widehat{n}(E,\lambda)\equiv\int\limits_0^\infty e^{-\lambda t}n(E,t)dt
$$
is the Laplace transform of the spectral function in the time
variable. The effect of the kinetic fluctuations is now taken into
account at the stage of the construction of equations for the
distributions $f(\mathbf{p}, t)$ or $n(E, t)$ by the inclusion of
additional terms containing differential and integral operators.
The theory of the acceleration of cosmic rays is under development
and this work is focused on the role of fractional derivatives in
this development.

\section{Fractional kinetic equations}

As was mentioned in [1], Eq. (1) is the diffusion limit of the
coordinate ($\mathbf{x}-CTRW$) model, where the trajectory of a
particle is a sequence of instantaneous jumps at random times by
random distances; the particle is at rest between these times. The
CTRW model with a finite velocity of free motion provides
qualitatively new results (see, e.g. [5]). When analyzing the
acceleration of cosmic rays in the first approximation, the
stochastic dynamics of the particle in the velocity (momentum or
energy) space is considered, and the boundedness of the coordinate
space is taken into account by correcting the mean age $\tau$ or
introducing additional coefficients characterizing the leakage of
the particles from the acceleration region or from the Galaxy as a
whole.

In the momentum space, the stationary position of the point that
does not coincide with the coordinate origin means the motion of
the particle with a constant momentum (velocity, energy). The
exponent $\beta$ characterizes now the tail part of the
distribution of the random interval duration $\Delta T$ between
the successive collisions of the moving particle
$Q(t)=\textsf{P}(\Delta T>t),\ t\to\infty$. For the
ultrarelativistic particle ($v\approx  c$), this interval is
proportional to its mean free path between the successive
collisions, so that the new $\beta$ values should correspond to
the $\alpha$ values in the preceding $\mathbf{x}-CTRW$ model. In
view of this circumstance, the consideration starts with the CTRW
system of Eqs. (2)-(4) from [1], where the coordinates
$\mathbf{r}$ are changed by the momenta $\mathbf{p}$, the momentum
change rate $\mathbf{v}$ is accepted to be infinite, and it is
assumed that the particle is injected by the source at the time
$t$ = 0 with the distribution function $f_0(\mathbf{p})$:
$$
f(\mathbf p, t)=\int \limits_0^t Q(t-t')F_{0\leftarrow
1}(\mathbf p, t')dt', \eqno(4a)
$$
$$
F_{0\leftarrow 1}(\mathbf p, t) = \int w(\mathbf p\leftarrow\mathbf
p')F_{1\leftarrow
0}(\mathbf p', t)d\mathbf{p}'  + f_0(\mathbf p) \delta
(t),\eqno(4b)
$$
$$
F_{1\leftarrow 0}(\mathbf p, t) = \int \limits_0^t q(t-t')
F_{0\leftarrow 1}(\mathbf p, t')dt'.\eqno(4c)
$$
The further transformation of the system is associated with the
specification of the distributions $q(t)$ and
$w(\mathbf{p}\leftarrow \mathbf{p}')$ distributions. The time is
usually (one can say, always) taken in the exponential form $Q(t)
= e^{-\mu t}, q(t)\equiv -dQ/dt = \mu e^{-\mu t}$ and therefore
the master equation begins with the first-order time derivative
$\partial f(\mathbf{p},t)/\partial t=\dots$. This means that the
process is assumed to be Markovian. However, the real distribution
of the time intervals between collisions is unknown. For example,
one can assume that it is of power law, $Q(t)\propto t^{-\alpha}$,
rather than exponential. This is in agreement with the
self-similar pattern of turbulent motions and with its power-type
laws. The hypothesis of the fractal character of the interstellar
medium [6] also provides power-law distributions. This concerns
the behavior of magnetic field lines in interstellar space. They
are usually represented as relatively smooth lines, which
somewhere rest at the magnetic traps, intersect each other,
sharply change their directions, and performing a "diffusion
dance" in time. The leading centers of particles moving along
spiral trajectories move along these lines. If these smooth
sections become invisible and chaotic patterns of the structure
become prevalent with the expansion of the field of view and the
corresponding decrease in the scale, then this is an
asymptotically homogeneous medium where the mean free paths can be
simulated by a usual exponential function. If the expanding field
of view includes increasing straight segments replacing the
segments becoming small due to a decrease in the scale so that the
structure remains (qualitatively) unchanged, this is a fractal
structure. Under these conditions, the exponential distribution of
mean free paths characteristic of a strongly mixing medium cannot
be expected, but can not be completely rejected as well. The best
compromise would be a family of distributions including both
exponential and power-law distributions. Fortunately, such a
family exists: it is a set of the functions
$$
Q_\alpha(t)=E_\alpha(-\mu t^\alpha),\quad \alpha\in(0,1],
$$
where $E_\alpha(z)=\sum\limits_{n=0}^\infty z^n/\Gamma(\alpha
n+1)$ are the Mittag-Leffler functions. The function $Q_\alpha(t)$
with $\alpha = 1$ is a usual exponential and with $\alpha < 1$ is
a fractional exponential having a power-law asymptotic behavior
$t^{-\alpha},\ t\to\infty$. The corresponding density
$q_\alpha(t)$ satisfies a fractional differential equation [7]; as
a result, the entire system of Eqs. (4a)-(4c) has the fractional
time-differential form
$$
_0\textsf{D}_t^\alpha f(\mathbf{p},t)=\mu \textsf{A}f(\mathbf{p},t)
+f_0(\mathbf{p})\delta_\alpha(t).\eqno(5)
$$
Here
$$
\textsf{A}f(\mathbf{p},t)=\int
w(\mathbf p\leftarrow\mathbf
{p}')f(\mathbf{p}',t)d\mathbf{p}' -f(\mathbf{p},t)\eqno(6)
$$
is the acceleration integral with transitions density $w(\mathbf
p\leftarrow\mathbf {p}')$. The time series of accelerating
collisions forms a \textit{fractional Poisson process of order}
$\alpha$ [7], with $\alpha\to1$ becoming  an ordinary Poisson
process which underlies classical kinetic equation (see (21.1)in
[8]). Investigations performed in [9] indicate a qualitatively new
property of this process: the average number of collisions
increases more slowly ($\propto t^{\alpha}$) than in the usual
case ($\propto t$) and relative fluctuations of the number of
collisions in the limit $t\to\infty$ do not disappear, but tend to
a limiting distribution depending on $\alpha$ (some sort of KNO
scaling).

The transition from the first time derivative to the fractional
derivative of the order $\alpha<1$ does not require solving a
fractional differential equation; it is more convenient to use the
relation between the solutions of Eq. (5) with $\alpha<1$ and
$\alpha=1$ [7],
$$
f_\alpha(\mathbf{p},t)=(t/\alpha)\int\limits_0^\infty f_1(\mathbf{p},\tau)
g_+(t\tau^{-1/\alpha};\alpha)
\tau^{-1/\alpha-1}d\tau.\eqno(7)
$$
Here $g_+(t;\alpha)$ is the one-sided stable (in the
L$\acute{e}$vy sense) probability density determined by the
Laplace transform
$$\int_0^\infty \exp(-\lambda
t)g_+(t;\alpha)dt=\exp(-\lambda^\alpha),$$ and $t,\ \tau$ ,
$\lambda$ are dimensionless variables. The Laplace transform of
Eq. (7) in time with the use of the above formulas for the spectra
provides the formula
$$
N_\alpha(E;\tau)=N_1(E;\tau^\alpha).\eqno(8)
$$
It presents the effect of the fractal dimension of the fractional
Poisson process of collisions $\alpha\in(0,1]$ on the energy
spectrum of cosmic rays: the spectrum $N_\alpha(E;\tau)$, formed
by an ensemble of particles with the mean life time $\tau$, which
are accelerated according to the fractional Poisson law of the
order $\alpha<1$, coincides with the spectrum of particles that
are accelerated by the usual Poisson process ($\alpha=1$), but
have the mean age $\tau$ (recall that the time is here
dimensionless and the injection spectra $f_0(E)$ are the same in
both problems). On the example of the Fermi spectrum, it is easy
to see that the efficiency of acceleration decreases (the spectrum
becomes steeper) with a decrease in the order of the process. The
fractal character of the spatial distribution of accelerating
regions naturally reduces the efficiency of acceleration.

\section{Fractional Fokker-Planck equations}

Similar to the classical case, the transition from the kinetic
equation to the Fokker–Planck equation is associated with the
transformation of the collision integral to a differential form by
expanding the integrand into a series in the momentum increment to
the second order terms. There are two variants of such an
expansion, which provide slightly different equations (see, e.g.,
[8]). The first variant implies the smallness of the
\textit{absolute value of the change in the momentum} $|\Delta
\mathbf{p}|=|\mathbf{p}-\mathbf{p}'|$, so that the momentum of the
incident particle only slightly changes in the magnitude and
direction in a single collision event (e.g., as in the case of the
collision of a heavy particle with a light one). The second
variant implies the smallness of the \textit{change in the
absolute value of the momentum} $\Delta
|\mathbf{p}|=|\mathbf{p}|-|\mathbf{p}'|$, whereas the change in
its direction is not small and can have a wide distribution up to
the isotropic one (in the case of the collision of a light
particle with a heavy one). Under the assumption of the isotropic
scattering, the fractional differential generalization of the
Fokker– Planck equation is obtained in the form
$$
_0\textsf{D}_t^\alpha f(p,t)=\triangle_\mathbf{p}(K(p)f(p,t))+ f_0(p)\delta_\alpha(t),\eqno(9)
$$
where
$$
K(p)=(\mu/2)\int(\Delta \mathbf{p})^2w(\Delta
\mathbf{p};p)d\Delta \mathbf{p}
$$
is the diffusivity in the momentum space. An energy analog of Eq.
(9) ($\alpha=1$) is well known in the physics of cosmic rays in
the form (see Eq. (14.2) in [10])
$$
_0\textsf{D}_t^\alpha n(E,t)=\frac{\partial[a_1(E) n(E,t)]}{\partial E}+
\frac{\partial^2[a_2(E) n(E,t)]}{\partial E^2}+n_0(E)\delta_\alpha(t).\eqno(10)
$$
At the same time, Eq.(9) with diffusion term
$$
\triangle_\mathbf{p}(K(p)f(p,t))=(\triangle_\mathbf{p} K(p))f(p,t)+2(\nabla_\mathbf{p} K(p))
\nabla_\mathbf{p} f(p,t)+K(p)\triangle_\mathbf{p} f(p,t)
$$
significantly differs from another diffusion type equation (см.
(9.57), [10])
$$
_0\textsf{D}^\alpha_t f(p,t)=\nabla_\mathbf{p} (K(p)\nabla_\mathbf{p} f(p,t))+
f_0(p)\delta_\alpha(t).\eqno(11)
$$
The difference is due to the fact that Eq. (9) is derived in frame
of the collision model, when the point presenting the particle
instantaneously moves to another, possibly far, geometric point,
violating the continuity of the trajectory in the momentum space,
whereas the dynamic derivation of Eq. (11) implies that the
trajectory in the momentum space is continuous and even
differentiable. The classical versions (with $\alpha=1$ ) of Eqs.
(9)-(11) underlie the standard set of mathematical tools
describing the fluctuation mechanisms of the acceleration of
cosmic rays; their solutions are well known [4, 10, 11]. A common
property of these models is the Gaussian character of the momentum
distributions, which is due to the assumption that the second
moment of the momentum acquired in an acceleration event, which
enters into the diffusion coefficient, is finite and the
acceleration rate is low. A change of the first time derivative in
this equation to its fractional analog of the order $\alpha\in(0,
1)$ does not lead to an increase in the efficiency of
acceleration; on the contrary, the acceleration rate in the
subdiffusion regime decreases further. It is also noteworthy that
the statement of some authors that diffusion at $\alpha > 1$ is
accelerated is erroneous. This statement is based on linguistic
intuition ("if $\alpha < 1$ means subdiffusion and $\alpha= 1$
corresponds normal diffusion, then $\alpha> 1$ should mean super
diffusion") and is erroneous, because the solution $f(\mathbf{p},
t)$ at $\alpha> 1$ loses its probability meaning: it is not
positively definite in this case. Thus, the transition to the
fractional time derivative (with the unchanged remaining,
diffusion part of the equation) as a method for enhancing the high
energy part of the spectrum is physically unpromising. At the same
time, the parameter $\alpha$ presents the existence of possible
correlations in the spatial distribution of the acceleration
regions and can be useful in this respect (recall that $\alpha =
1$ corresponds to the uniform Poisson distribution of such
regions, which do not correlate with each other). This parameter
can be kept and the efficiency of the multiple acceleration
mechanism can be increased only by modifying another operator of
the equation, more precisely, by returning from the differential
form of the acceleration operator, which describes a continuous
slow collection of the energy, to the integral form describing
acceleration as a sequence of events with large instantaneous (in
terms of the considered "galactic" time scales) changes in the
momenta. For this reason, it seems appropriate to change the
momentum Laplacian to its fractional analog, because the
fractional Laplacian contains a momentum integral operator with
inverse power law kernel, which can ensure a high acceleration
rate. Let us consider this variant in more detail.

The losses of the energy of a fast charged particle in a medium,
which is described by equations similar to Eq. (10) (naturally,
with $\alpha = 1$), cannot be larger than its initial energy and
all of the moments of the energy loss are finite. In the problem
of acceleration, there is no such definite limit of the energy
increase; this fact is an additional reason to study the region
with infinite dispersion, which attracts an increasing interest of
researchers of anomalous diffusion processes. In this case,
informative (in the asymptotic sense) results are obtained only
when infinite dispersion is due to the power-law distributions:
$$
\int\limits_{|\Delta \mathbf{p}|>p}w(\Delta
\mathbf{p};\mathbf{p}')d\Delta \mathbf{p}\propto p^{-\gamma},\
p\to\infty .\eqno(12)
$$
If $\gamma > 2$, the second moment is finite and this corresponds
to the classical diffusion region. If $\gamma < 2$, the second
moment of the increment is infinite and this corresponds to the
model of additive Levy flights. In this case, the equations for
the momentum and energy distributions f(p, t) and n(E, t),
respectively, following from the asymptotic CTRW analysis,
$$
_0\textsf{D}_t^\alpha f(p,t)=-K(-\triangle_\mathbf{p})^{\nu/2}
f(p,t)+f_0(p)\delta_\alpha(t),\eqno(13)
$$
and
$$
_0\textsf{D}_t^\alpha n(E,t)=\left\{
                               \begin{array}{ll}
                                 \partial^\nu[a_\nu n(E,t)]/{\partial E^\nu}+
n_0(E)\delta_\alpha(t), & \hbox{$0<\nu<1$;} \\
                                 \partial[a_1 n(E,t)]/\partial E+\partial^\nu[a_\nu n(E,t)]/
\partial E^\nu+n_0(E)\delta_\alpha(t), & \hbox{$1<\nu<2$.}
                               \end{array}
                             \right.
\eqno(14)
$$
seem to be reasonable. Here
$$
\nu=\left\{
      \begin{array}{ll}
        \gamma, & \hbox{$\gamma\leq 2$;} \\
        2, & \hbox{$\gamma>2$,}
      \end{array}
    \right.
$$
$\partial^\nu/\partial E^\nu$ is the symbol closer to the ordinary
notation of the fractional differential operator and $K,\ a_1$ and
$a_\nu$ are the constant coefficients. The constancy of these
coefficients is very significant for the derivation of the
equations. Fractional differential equations are usually derived
with the use of integral transformations whose applicability
requires the constancy of the coefficients. It would be incorrect
to derive, e.g., Eq. (13) in such a manner and, then, to place the
variable diffusion coefficient K(p) in front of the fractional
Laplacian (this is obvious even on the example of Eqs. (9)– (11)
with the integer Laplacian).

Equation (14) is clearly similar to the $\nu$th term approximation
of the expansion of a function that is zero at the reference point
and it would seem that omitting the next (divergent!) term
introduces an infinite error, but this is not the case. The reason
is that the Taylor formula rather than the infinite Taylor series
is used and the series expansion can be continued only until
derivatives exist; after that, the remainder term, which is always
finite, should be written. If the next derivative existed, its
inclusion in the continued expansion would mean an approximation
of the omitted term. If this derivative does not exist, it cannot
be used to approximate the omitted term and one should return to
the initial point, where the derivative of the preceding order is
used for approximation. The diffusion packet, which is described
by Eq. (13) and propagates from the origin of the momentum
coordinates, has the form of the three-dimensional isotropic
fractional stable distribution $\psi_3^{(\nu,\alpha)},\
\nu\in(0,2],\ \alpha\in(0,1]$, smearing proportionally to
$t^{\alpha/\nu}$:
$$
f(p,t)=
(Kt^\alpha)^{-3/\nu}\psi_3^{(\nu,\alpha)}((Kt^\alpha)^{-1/\nu}p).
$$
Tails (or, more romantically, wings) of this distribution have a
power-law form with the exponent $\nu$. Physically, this means a
peculiar leading effect: one of independent terms in the sum
$\Delta \mathbf{p}_1+\Delta \mathbf{p}_2+\dots+\Delta
\mathbf{p}_n$ is always outstanding and compared in magnitude with
the reminder. This leading effect disappears at $\nu = 2$, when
the distribution becomes Gaussian (sub-Gaussian). As a result, the
spectrum at $\nu < 2$ has the form
$$
N_1(E)dE\propto E^{-\nu-1}dE,
$$
which is similar to the Fermi formula. The main difference is that
the exponent $\nu$ in this case is independent of the age of the
detected particles and is completely determined by the
acceleration mechanism in an individual local event (collision).
For this reason, the fractal character of the spatial distribution
of accelerating regions also does not affect the slope of the
resulting spectrum.

\section{Integro-fractionally-differential model}

A drawback of the model with the fractional momentum Laplacian is
that the increments of the momentum in the acceleration event are
independent of the momentum of the particle involved in the
interaction, whereas in the Fermi model and in its later variants,
the increments of the energy (and, therefore, the momentum) are on
average proportional to the energy (momentum) of the particle
before the interaction. In this case, the energy of the
accelerated particle is expressed in terms of the product of
independent random variables, rather than their sum. This model is
called \textit{multiplicative walk} in order to distinguish it
from the \textit{additive walk} model considered above. The
increment of the momentum in the multiplicative model is
proportional (in the statistical case) to the absolute value of
the momentum $p'$ of the particle coming into interaction,
$$
\Delta\mathbf{p}=p'\mathbf{q},\quad \int\limits_{|\Delta
\mathbf{p}|>p}w(\Delta \mathbf{p};\mathbf{p}')d\Delta
\mathbf{p}\propto (p/p')^{-\gamma},\ p\to\infty .\eqno(15)
$$
Under the assumption that the distribution of the proportionality
vector $\mathbf{q}$ is independent of $\mathbf{p}'$ and isotropic,
$$W(\mathbf{q};\mathbf{p}')d\mathbf{q}=(1/2)V(q)dqd\xi,\
\xi=\cos({\mathbf{q},\mathbf{p})}$$, kinetic equation (5) can be
modified to the form
$$
_0\textsf{D}_t^\alpha f(p,t)=\mu\left\{\int
\limits_{-1}^1\frac{d\xi}{2}\int\limits_0^\infty
V(q)f\left(p\bigg/\sqrt{1+2\xi q+q^2},t\right)\bigg/
\left(\sqrt{1+2\xi
q+q^2}\right)^3dq-f(p,t)\right\}+f_0(p)\delta_\alpha(t),\eqno(16)
$$
representing a new model of a distributed reacceleration, more
precisely, a new modification of the model proposed in [12, 13].
In order to make this model be closer to real processes of
reacceleration, e.g., in the case of the intersection of shock
fronts in the remnants of supernovae, we assume that [13]
$$
V(q)=\gamma q^{-\gamma-1},\quad \gamma>1.
$$
The resulting model can be called \textit{multiplicative Levy
flights}.

Let us consider the equation for the spectral function in two
extreme cases. In the first case, $\gamma
>2$, the second moment of the momentum increment proportional to $E^2$ exists, and this case
corresponds to the classical diffusion with variable coefficients:
$$ _0\textsf{D}_t^\alpha n(E,t)=\frac{\partial[a_1E n(E,t)]}{\partial E}+
\frac{\partial^2[a_2E^2 n(E,t)]}{\partial E^2}+n_0(E)\delta_\alpha(t).
$$
In the second case, we suppose $\gamma << 2$, so that only the
term $q^2$ may be retained in the radicand in Eq. (16):
$$
_0\textsf{D}_t^\alpha n(E,t)=\mu\left\{
\int\limits_1^\infty\gamma q^{-\gamma-1}n(E/q,t)dq/q-n(E,t)\right\}+
n_0(E)\delta_\alpha(t).    \eqno(17)
$$
I do not claim that this approximation is very good, but namely
this acceleration operator was used in [13] for some calculations.
Solving Eq. (17) with the method of the Mellin– Laplace transforms
and using Eqs. (3) and (8), the following expression is obtained
for the case of the mono-energetic source ($n(E)=\delta(E-E_0)$):
$$
N_\alpha(E;\tau)=\frac{\mu\tau^\alpha\gamma}{(1+\mu\tau^\alpha)^2}\left(\frac{E}{E_0}\right)^
{-1-\gamma/(1+\mu \tau^\alpha)}\frac{1}{E_0}.\eqno(18)
$$
Although Eq. (18) was derived under the assumptions that strongly
simplify the real situation and has a qualitative sense, it
compactly presents the effect of \textit{all three sources of
fluctuation acceleration}: fluctuations of the age of the particle
(parameter $\tau$), fluctuations of the number of the acceleration
events ($\alpha$ и $\mu$), and fluctuations of the energy acquired
in a single event ($\gamma$). Representing the scaling parameter
$\mu$ in the form $\mu=\tau_{\rm A}^\alpha$, where $\tau_{\rm A}$
is the characteristic time interval between the events of the
acceleration of particles in the remnants of various supernovae
(recall that $\tau$ is the mean lifetime with respect to nuclear
collisions), the absolute value of the exponent of the integral
spectrum can be written in the more clear form
$\gamma'=1+\gamma/[1+(\tau/\tau_{\rm A})^\alpha]$. At $\alpha=1,$
and $\mu\tau\gg 1$, we arrive at the Fermi formula (2) with $a =
\mu/\gamma$.

\section{Conclusions}

Taking into account that all three above-considered mechanisms
contribute to the acceleration (more precisely, reacceleration) of
cosmic rays, this process can be analyzed using the
integro-fractional-differential equation
$$
_0\textsf{D}_t^\alpha f(p,t)=\mu_0\nabla_\mathbf{p} (K_0(p)\nabla_\mathbf{p} f(p,t))
+
\mu_1\triangle_\mathbf{p}(K_1(p)f(p,t))
-\mu_2K_2(-\triangle_\mathbf{p})^{\nu/2}
f(p,t)+
$$
$$
\mu_3\left\{\int \limits_{-1}^1\frac{d\xi}{2}\int\limits_0^\infty
V(q)f\left(p\bigg/\sqrt{1+2\xi q+q^2},t\right)\bigg/
\left(\sqrt{1+2\xi
q+q^2}\right)^3dq-f(p,t)\right\}+f_0(p)\delta_\alpha(t).\eqno(19)
$$
The terms with the coefficients $\mu_0,\ \mu_1,\ \mu_2$ and
$\mu_3$ successively present the contributions from the processes
of continuous acceleration (such as fluctuations of the momentum
of the particle in turbulent plasma [11, 14]), collision
accelerations with a finite second moment (the model of walking
magnetized clouds with a limited distribution of their
velocities), additive Levy-type collision accelerations given by
Eq. (12) (the same with the power-law distribution of the
velocities of the clouds), and multiplicative Levy-type collision
accelerations given by Eq. (15). Recall again that the last
operator describing, in particular, acceleration at shock waves in
the remnants of supernovae does not reduced to the fractional
differential form, but holds in the integral form. Equation (19)
should be supplemented by terms presenting energy losses, exist of
particles from the Galaxy, and nuclear interactions
(fragmentation).

However, only dominant terms can be retained in Eq. (19) applied
to particular problems, as is usually done.

\section{Acknowlegments}
I am grateful to Prof. M.I. Panasyuk and all the participants of
his seminar at the Skobeltsyn Institute of Nuclear Physics, Moscow
State University, for the discussion of some problems of this work
in February 2010, and to V.S. Ptuskin and L.G. Sveshnikova for
valuable consultations. This work was supported by the Russian
Foundation for Basic Research, project no. 10-01-00608.

\end{document}